\documentclass[%
showpacs,
twocolumn,
 amsmath,amssymb,
]{revtex4}

\usepackage{graphicx}
\usepackage{dcolumn}
\usepackage{bm}

\newcommand{\Ra}{\mbox{Ra}}
\newcommand{\Nu}{\mbox{Nu}}

\def\u{\mbox{\boldmath $u$}}

\def\i{\mbox{\boldmath$\hat{i}$}}
\def\j{\mbox{\boldmath$\hat{j}$}}

\begin{document}

\preprint{AIP/123-QED}

\title{``Ultimate state'' of two-dimensional Rayleigh-B\'enard convection\\
between free-slip fixed-temperature boundaries}

\author{Jared P. Whitehead$^{1,}$}
\email{jaredwh@umich.edu}
\author{Charles R. Doering$^{1,2,3,}$}
\email{doering@umich.edu}
\affiliation{ 
{$^1$Department of Mathematics, University of Michigan, Ann Arbor, MI 48109-1034} \\
{$^2$Department of Physics, University of Michigan, Ann Arbor, MI 48109-1040} \\
{$^3$Center for the Study of Complex Systems, University of Michigan, Ann Arbor, MI 48109-1107}
}


\date{\today}

\begin{abstract}
Rigorous upper limits on the vertical heat transport in two dimensional Rayleigh-B\'enard convection between stress-free isothermal boundaries are derived from the Boussinesq approximation  of the Navier-Stokes equations.
The Nusselt number $\Nu$ is bounded in terms of the Rayleigh number $\Ra$ according to $\Nu \leq 0.2295 \, Ra^{5/12}$ uniformly in the Prandtl number $\Pr$.
This Nusselt number scaling challenges some theoretical arguments regarding the asymptotic high Rayleigh number heat transport by turbulent convection.
\end{abstract}

\maketitle

Rayleigh-B\'enard convection is the buoyancy-driven flow of a fluid heated from below and cooled from above.  
It is important for a variety of systems in the engineering, geophysical, and astrophysical sciences, and it has long served as a fundamental paradigm of nonlinear science, chaos, and pattern formation.
Indeed, the Boussinesq approximation to the Navier-Stokes equations with the boundary conditions analyzed in this paper was Rayleigh's original model for calculating conditions for onset  \cite{Rayleigh16}, it is the basis of the Lorenz equations \cite{Lorenz63}, and it formed the foundation of developments in the modern mathematical theory of amplitude \cite{MV58} and modulation \cite{NewellWhitehead69} equations.
Most recently Rayleigh-B\'enard convection has been the focus of a large body of experimental, computational, theoretical, and mathematical research aimed at characterizing the fully turbulent dynamics for application in geophysical and astrophysical regimes \cite{AhGrLo2009}.

Convective fluid flow increases vertical heat transport beyond  the purely conductive flux.
The dimensionless enhancement factor, the Nusselt number $\Nu$, is both of fundamental interest for applications and the natural and widely recognized measure of the intensity and effectiveness of the motion.
The most basic question for Rayleigh-B\'enard convection is the dependence of $\Nu$ on (i) the strength of the thermal forcing, commonly expressed in terms of a dimensionless Rayleigh number $\Ra$, (ii) the material properties of the fluid, which within the Boussinesq approximation is set by the dimensionless Prandtl number $\Pr$, the ratio of the fluid's momentum and thermal diffusion coefficients, (iii) the geometry, typically the aspect ratio of the container, and (iv) the boundary conditions.
The connection between these variables is generally complex and often not even unique, but in the ``ultimate''  high Rayleigh number regime when the flow is turbulent, the presumed functional relation between the $\Nu$, $\Pr$, and $\Ra$ is $\Nu \sim \Pr^{\gamma} \Ra^{\beta}$.

Experiments and simulations with $\Pr = {\cal O}(1)$ and {\it no-slip} boundary conditions suggest a scaling exponent $0.27 \lesssim \beta \lesssim 0.40$ at the highest available $\Ra$ \cite{AhGrLo2009,RGKS2010}.
Various theories suggest (modulo possible logarithmic corrections) that $\Nu \sim \Pr^{1/2} \Ra^{1/2}$ as $\Ra \rightarrow \infty$ \cite{Kraichnan62,Spiegel71,GrLo2000}.
Rigorous analyses of the Boussinesq model with no-slip velocity  and isothermal (fixed temperature) \cite{Ho1963,DoCo1996} or fixed heat flux \cite{OtWiWoDo2002} or mixed temperature \cite{Wittenberg2010} boundary conditions yield upper bounds of the form $\Nu \le c \, \Ra^{1/2}$ with prefactors $0 < c < \infty$ independent of $\Pr$, so $\beta=\frac{1}{2}$ and $\gamma=\frac{1}{2}$ cannot both hold for very large $\Pr$.
The $\Nu$-$\Ra$ relation is certainly different at $\Pr = \infty$ where theory suggests  \cite{Malkus1954} and analysis proves \cite{DoOtRe2006} (modulo possible logarithmic corrections) that $\Nu \lesssim \Ra^{1/3}$.

Two dimensional Rayleigh-B\'enard convection displays many of the physical and turbulent transport features of three dimensional convection and has long been utilized as a test-bed for theoretical concepts \cite{DWRC1990,JoDo2009}.
The effect of free-slip (no-stress) velocity boundary conditions on developed turbulent convection has largely been unexplored although we note that the rigorous scaling bound reported here was anticipated by recent numerical and perturbative investigations of transport limits for finite \cite{Otero2002} and infinite \cite{IeWo2001,IeKePl2006} Prandtl numbers.
This letter bridges that gap with a proof that $\Nu \leq 0.2295 \, Ra^{5/12}$ uniformly in $0 < \Pr \le \infty$ for the Boussinesq model in two spatial dimensions with fixed temperature and free-slip boundaries.
This result refutes predictions of a $\Nu \sim Ra^{1/2}$ ultimate regime insofar as the theoretical arguments do not refer specifically to the boundary conditions or the spatial dimension.
This issue is discussed further in the conclusion section at the end of the paper.
Meanwhile the proof of the bound is presented in sufficient detail immediately below for motivated readers to reproduce the calculation in its entirety.
The key new idea used to derive the result emerged from intuition developed in numerical studies of upper bounds \cite{IeWo2001, Otero2002}: implement and exploit  the bulk averaged enstrophy balance available for two-dimensional flows with free-slip boundaries to decrease the upper bound.

\begin{figure}
\includegraphics[scale=0.3]{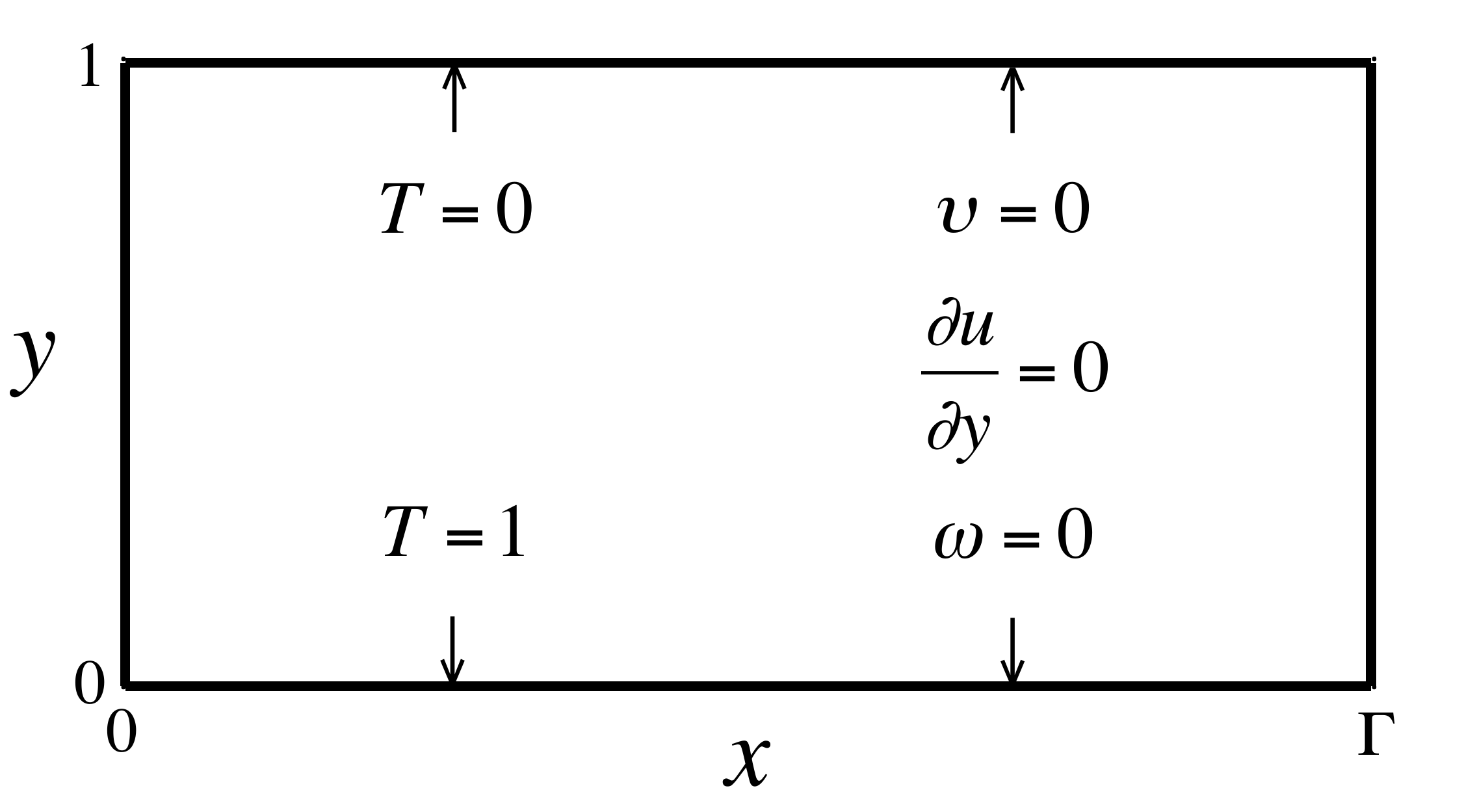}
\caption{Geometry for the 2d stress-free convection problem.  Boundary conditions for $T$, $u$, $v$, and the vorticity $\omega$ at the isothermal no-slip vertical boundaries are shown.
All these variables as well as the pressure $p$ are periodic in the horizontal direction with period $\Gamma$.}
\label{fig:setup}
\end{figure}

The dimensionless equations of motion for the Boussinesq approximation are
\begin{eqnarray}\label{eq:mom}
\frac{1}{\Pr}\left(\frac{\partial \u}{\partial t} + \u \cdot \nabla \u\right) + \nabla p &=& \nabla^2 \u + \Ra \, \j \, T, \\ 
\nabla \cdot \u &=& 0, \label{eq:div_free} \\ 
\frac{\partial T}{\partial t} + \u \cdot \nabla T &=& \nabla^2 T, \label{eq:temp}
\end{eqnarray}
where the Prandtl number $\Pr = \nu / \kappa$ is the ratio of the fluid's kinematic viscosity $\nu$ to its thermal diffusivity $\kappa$, and the Rayleigh number $\Ra = g \, \alpha \, \Delta T \, h^3 / \nu \, \kappa$ where $g$ is the acceleration of gravity, $\alpha$ is the fluid's thermal expansion coefficient, and $\Delta T$ is the imposed temperature drop across the layer of thickness $h$.
Lengths are measured in units of $h$, time in units of $h^2/\kappa$, and temperature in units of $\Delta T$.
The velocity vector field $\u(x,y,t) = \i u(x,y,t) + \j v(x,y,t)$ satisfies no-penetration and free-slip (stress-free) boundary conditions, and the temperature field $T(x,y,t)$ is isothermal on the vertical boundaries at $y=0$ and $y=1$ as shown in Fig. \ref{fig:setup}.
All dependent variables, $u$, $v$, $T$, and the pressure field $p(x,y,t)$, are periodic in the horizontal direction $x$ with period $\Gamma$ (the aspect ratio).

Taking the curl of \eqref{eq:mom} one obtains the evolution equation for the scalar vorticity $\omega = \partial v / \partial x - \partial u/ \partial y$,
\begin{equation}\label{eq:vort}
\frac{1}{\Pr}\left(\frac{\partial\omega}{\partial t} + \u \cdot \nabla \omega\right) = \nabla^2 \omega + \Ra \, \frac{\partial T}{\partial x}.
\end{equation}
The boundary conditions on $u$ and $v$ imply that $\omega = 0$ on the vertical boundaries at $y=0$ and $y=1$.

The goal of the analysis is to use the equations of motion to derive upper bounds on the Nusselt number defined as $\Nu = 1 + \langle v T\rangle$, where $\langle\cdot\rangle$ represents the spatial and long time average, in terms of $\Ra$, $\Pr$, and $\Gamma$.
Toward this end we utilize the background method \cite{DoCo1992}, a mathematical device introduced by Hopf to establish the existence of weak solutions to the Navier-Stokes equations in bounded domains \cite{Ho1941}.
For convection problems the background method involves decomposing the temperature field into a background profile $\tau(y)$ which satisfies the vertical boundary conditions ($\tau(0)=1$ and $\tau(1)=0$) and a perturbation term $\theta(x,y,t)$ satisfying corresponding homogeneous boundary conditions ($\theta(x,0,t) = 0 = \theta(x,1,t)$) so that $T(x,y,t) = \tau(y) + \theta(x,y,t)$ \cite{DoCo1996}.
Implementing this decomposition the temperature equation \eqref{eq:temp} implies
\begin{eqnarray}
\frac{\partial \theta}{\partial t} + \u \cdot \nabla \theta = \nabla^2 \theta + \tau''(y) - v  \tau'(y).
\end{eqnarray}
Then the equations of motion together with the boundary conditions and the background decomposition imply
\begin{eqnarray}\label{eq:energy_mom}
\frac{1}{2\Pr}\frac{d}{dt}\|\u\|_2^2 &=& -\|\omega\|_2^2 + \Ra\int v\, \theta \, dxdy  \\ \label{eq:energy_omega}
\frac{1}{2\Pr}\frac{d}{dt}\|\omega\|_2^2 &=& -\|\nabla\omega\|_2^2 + \Ra\int \omega \, \frac{\partial \theta}{\partial x} \, dxdy \\ \label{eq:energy_theta}
\frac{1}{2}\frac{d}{dt}\|\theta\|_2^2 &=& -\|\nabla\theta\|_2^2 - \int \left[ \tau' \frac{\partial \theta}{\partial y} +  \tau'  v \theta \right] dx dy \\ \label{eq:energy_T} 
\|\nabla T\|_2^2 &=& \|\nabla\theta\|_2^2 + 2 \int \tau'  \frac{\partial \theta}{\partial y} dx dy + \| \tau' \|_2^2 
\end{eqnarray}
where $\|\cdot\|_2$ is the $L^2$ norm on the spatial domain and the elementary identity $\|\nabla \u\|_2^2 = \|\omega\|_2^2$ was used in (\ref{eq:energy_mom}).

It is well-known that the equations of motion imply $\Nu = \langle |\nabla T|^2\rangle$  \cite{Ho1963,DoCo1996}.
Thus, given coefficients $a$ and $b$ with precise values to be determined, combining (\ref{eq:energy_mom}-\ref{eq:energy_T}) according to
\begin{equation}
\frac{b}{\Ra} \times \eqref{eq:energy_mom} + \frac{a}{\Ra^{3/2}} \times \eqref{eq:energy_omega} + 2 \times \eqref{eq:energy_theta} + \eqref{eq:energy_T} ,
\end{equation}
applying the long time average---remarking that it can be shown within the background method that the time averages of the time derivatives vanish \cite{DoCo1992,DoCo1996}---and dividing by $\Gamma$, the Nusselt  number is expressed
\begin{equation}\label{eq:Nu_bound}
\Nu = \frac{1}{1-b}\left(\int_0^1\tau'(y)^2 dy - b\right) - \frac{1}{1-b} \, \mathcal{Q}
\end{equation}
where
\begin{eqnarray}
\mathcal{Q} &=& \left\langle |\nabla\theta|^2 \ + \ \frac{a}{\Ra^{3/2}} |\nabla\omega|^2 \ + \ \frac{b}{\Ra} |\omega|^2 \right. \nonumber \\
&+& 2 \left. \tau' \, v \, \theta \ + \ \frac{a}{\Ra^{1/2}} \omega \, \frac{\partial \theta}{\partial x} \right\rangle.
\end{eqnarray}

\begin{figure}
\includegraphics[scale=0.3]{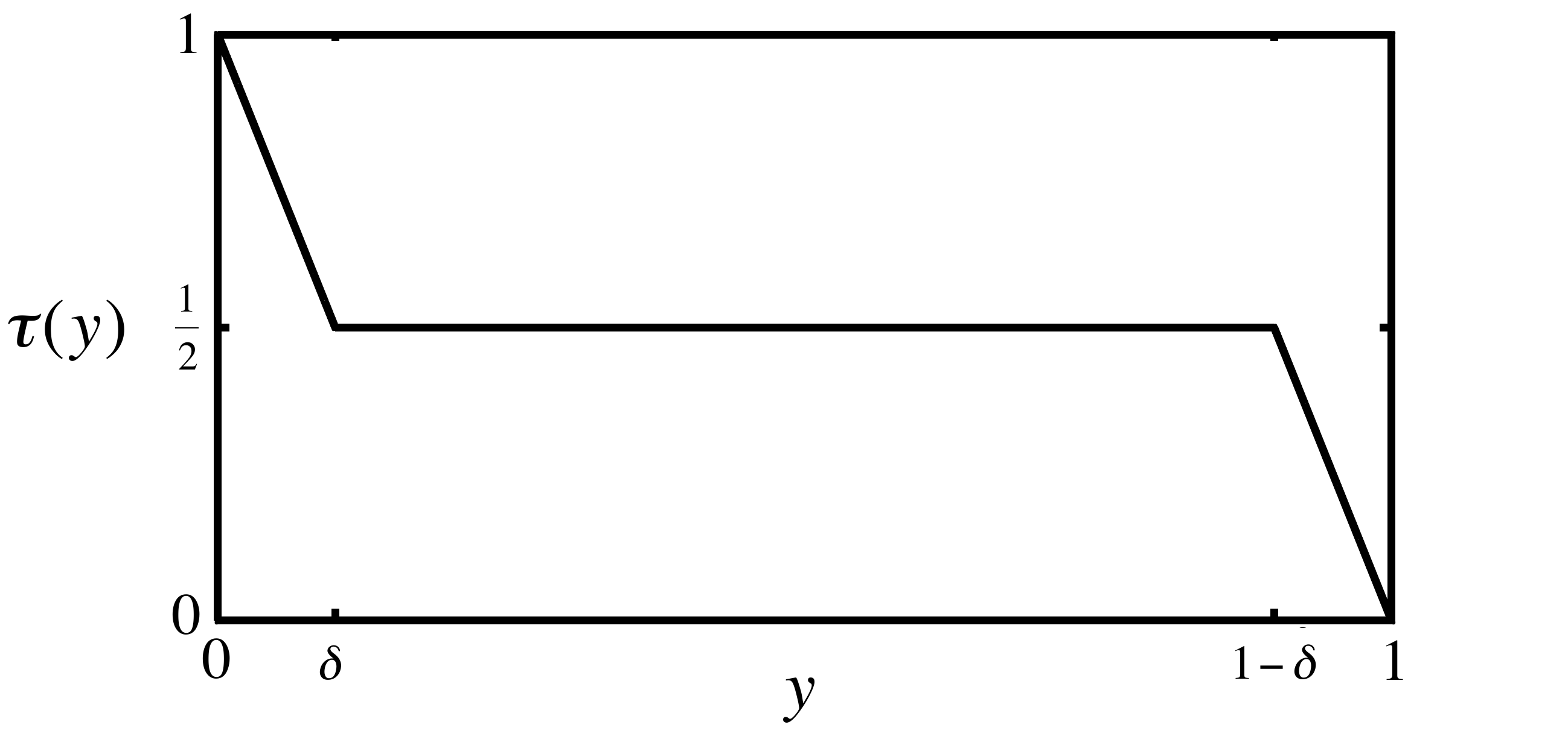}
\caption{Background profile with boundary layers of thickness $0 < \delta \le \frac{1}{2}$ in which $\tau'(y)=-\frac{1}{2\delta}$;
$\tau'(y) \equiv 0$ for $\delta < y < 1-\delta$.}
\label{fig:tau}
\end{figure}

Hence {\it if} we can choose  the background profile $\tau(y)$ and coefficients $a>0$ and $0<b<1$ so that $\mathcal{Q} \ge 0$ for all relevant $\theta$, $\omega$ and $v$, {\it then} the first term on the right hand side of \eqref{eq:Nu_bound} is an upper bound on $\Nu$.
For the problem at hand we may use the piece-wise linear profile shown in Fig. \ref{fig:tau} where the thickness $\delta$ of the ``boundary layers''  is to be determined as a function of $\Ra$ to satisfy $\mathcal{Q} \ge 0$. 
With this choice of $\tau(y)$ the bound will be
\begin{equation}\label{eq:Nu_delta_bound}
\Nu \leq \frac{1}{2\delta(1-b)}-\frac{b}{1-b}.
\end{equation}

Applying the horizontal Fourier transform and introducing the shorthand $D = \frac{d}{dy}$, it is evident that positivity of $\mathcal{Q}$ is equivalent to the positivity of
\begin{eqnarray} \nonumber
\mathcal{Q}_k \ = \ \|D\hat{\theta}_k\|^2+k^2\|\hat{\theta}_k\|^2 &+& \frac{a}{\Ra^{3/2}}\|D \hat{\omega}_k\|^2\\ 
+ \frac{a}{\Ra^{3/2}}k^2\|  \hat{\omega}_k\|^2 &+& \frac{b}{\Ra}\|  \hat{\omega}_k\|^2\\ \nonumber
+ \ \text{Re} \left\{2\int_0^1 \tau' \, \hat{v}_k \, \hat{\theta}^*_k dy \right. &-& \left. \frac{a i k}{\Ra^{1/2}}\int_0^1  \hat{\omega}_k \, \hat{\theta}^*_k dy \right\}
\end{eqnarray}
for each horizontal wavenumber $k$ where $\|\cdot\|$ is now the $L^2$ norm on complex valued functions of $y \in [0,1]$ and $\text{Re}\left\{ \cdot \right\}$ indicates the real part of a complex quantity.
The Cauchy-Schwarz and Young inequalities imply
\begin{equation}
\left|\frac{a \, i \, k}{\Ra^{1/2}}\int_0^1 \hat{\omega}_k \hat{\theta}^*_k dy \right| \ \leq \ \frac{a^2}{4\Ra}\|\ \hat{\omega}_k\|^2 + k^2\| \hat{\theta}_k\|^2
\end{equation}
so dropping the manifestly non-negative term $\|D \hat{\omega}_k\|^2$,
\begin{eqnarray} \nonumber
\mathcal{Q}_k \ \ge \ \|D \hat{\theta}_k\|^2 + \left[\frac{ak^2}{\Ra^{3/2}} \right. &+& \left. \frac{1}{\Ra}\left(b-\frac{a^2}{4}\right)\right]\|\ \hat{\omega}_k\|^2\\
- \frac{1}{\delta} \, \text{Re} \left\{ \int_0^\delta \hat{v}_k(y) \hat{\theta}_k^*(y) dy \right. &+& \left. \int_{1-\delta}^1 \hat{v}_k(y) \hat{\theta}_k^*(y) dy \right\}.
\end{eqnarray}
Restricting $a^2 < 4 b$, the task is to dominate the indefinite boundary layer integrals by the positive definite terms.

The Fourier coefficients of the vertical velocity and vorticity (suppressing the time dependence) are related by
\begin{equation}\label{eq:v_omega}
i k\ \hat{\omega}_k(y) = D^2 \hat{v}_k(y) - k^2 \hat{v}_k(y).
\end{equation}
Integrating the modulus squared of both sides with a simple integration by parts implies
\begin{equation}\label{eq:l2_omega1}
k^2\|\ \hat{\omega}_k\|_2^2 = \|D^2v_k\|^2 + 2k^2\|Dv_k\|^2 + k^4\|v_k\|^2.
\end{equation}
On the other hand, integration by parts and the Cauchy-Schwarz and Young inequalities yield
\begin{equation}
\frac{2}{3}k^2\|D \hat{v}_k\|^2 \ \leq \ \frac{1}{9}\|D^2 \hat{v}_k\|_2^2 + k^4\| \hat{v}_k\|^2 \label{Ted}
\end{equation}
so that, combining \eqref{eq:l2_omega1} and \eqref{Ted}, 
\begin{equation}
k^2\| \hat{\omega}_k \|_2^2 \geq \frac{8}{9}\|D^2 \hat{v}_k \|^2 + \frac{8}{3}k^2\|D \hat{v}_k \|^2.
\end{equation}
Boundary conditions on $\hat{v}_k(y)$ dictate that
\begin{equation}
\int_0^1 \text {Re} \left\{ D \hat{v}_k(y) \right\} dy = \left. \text {Re} \left\{\hat{v}_k(y) \right\} \right|_{z=0}^{z=1} = 0
\end{equation}
so $\exists \ y_0 \in (0,1)$ such that $\text {Re} \left\{ D \hat{v}_k(y_0) \right\}  = 0$.
The fundamental theorem of calculus followed by application of the Cauchy-Schwarz and Young inequalities imply
\begin{eqnarray} \nonumber
&\left( \text {Re} \left\{ D \hat{v}_k(y) \right\} \right)^2 = 2\int_{y_0}^y \text {Re} \left\{ D^2 \hat{v}_k(y') \right\} \text {Re} \left\{ D \hat{v}_k(y') \right\} dy' \\
&\leq \  \frac{\sqrt{27}}{8k}\left(\frac{8}{9}\| \text {Re}\left\{ D^2 \hat{v}_k \right\} \|^2 \right. + \left. \frac{8}{3}k^2 \| \text {Re}\left\{ D \hat{v}_k \right\} \|^2 \right).
\end{eqnarray}
A similar pointwise bound holds for the imaginary part of $D \hat{v}_k(y)$ so its modulus squared satisfies
\begin{eqnarray} \nonumber
&| D \hat{v}_k(y) |^2  \leq \  \frac{\sqrt{27}}{8k}\left(\frac{8}{9}\| D^2 \hat{v}_k \|^2 \right. + \left. \frac{8}{3}k^2\|  D \hat{v}_k \|^2\right) \\
&\leq \frac{3^{3/2}}{8} k \|\ \hat{\omega}_k\|^2.
\end{eqnarray}
Thus, integrating $D \hat{v}_k$ from $0$ to $y$ and applying H\"older's inequality, it is evident that 
\begin{equation}
|\hat{v}_k(y)| \leq \frac{3^{3/4}}{2^{3/2}} \, k^{1/2} \, y \, \|\ \hat{\omega}_k\|.   \label{eq:final_estimate1}
\end{equation}
Likewise, integrating $D \hat{v}_k$ from $1-y$ to $1$,
\begin{equation}
|\hat{v}_k(y)| \leq \frac{3^{3/4}}{2^{3/2}} \, k^{1/2} \, (1-y) \, \|\ \hat{\omega}_k\|.   \label{eq:final_estimate2}
\end{equation}

Because $ \hat{\theta}_k(y)$ vanishes at $y=0$ and $1$, applications of the fundamental theorem of calculus and Cauchy-Schwarz inequality yield the pointwise bounds
\begin{equation}\label{eq:theta_estimate1}
| \hat{\theta}_k(y)| \leq y^{1/2} \left(\int_0^{1/2} |D \hat{\theta}_k(y')|^2 dy' \right)^{1/2}
\end{equation}
for $0 \le y \le 1/2$ and, for $1/2 \le y \le 1$,
\begin{equation}\label{eq:theta_estimate2}
 | \hat{\theta}_k(y)| \leq (1-y)^{1/2} \left(\int_{1/2}^1 |D \hat{\theta}_k(y')|^2 dy' \right)^{1/2}.
\end{equation}
Using (\ref{eq:final_estimate1} - \ref{eq:theta_estimate2}), we conclude
\begin{eqnarray} \nonumber
& \frac{1}{\delta}\left| \int_0^\delta \hat{v}_k(y) \hat{\theta}_k^*(y) dy \right. + \left. \int_{1-\delta}^1 \hat{v}_k(y) \hat{\theta}_k^*(y) dy \, \right| \  \le \\ 
& \leq \ \frac{3^{3/2}}{5^2\cdot 2^3} \, k \, \delta^3 \, \|\ \hat{\omega}_k\|^2 + \|D \hat{\theta}_k\|^2.
\end{eqnarray}

Hence $\mathcal{Q}_k \geq 0$ is guaranteed by a $\delta$ small enough that
\begin{eqnarray}
\frac{ak^2}{\Ra^{3/2}} + \frac{1}{\Ra}\left(b-\frac{a^2}{4}\right) - \frac{3^{3/2}k}{5^2\cdot 2^3}\delta^3 \geq 0. \label{eq:delta_abk}
\end{eqnarray}
Inserting $a=\frac{2}{\sqrt{15}}$ and $b=\frac{1}{5}$ into \eqref{eq:delta_abk}---chosen to minimize the prefactor in the bound---and minimizing the suitable $\delta$ over $k$, this is satisfied by choosing $\delta = \frac{2^{5/3}\cdot 5^{5/12}}{3^{3/4}} \, Ra^{-5/12}$ where $k = \frac{1}{3^{1/4}\cdot 5^{1/4}} \, \Ra^{1/4}$ is the minimizing wavenumber.
Inserting these $\delta$ and $b$ into \eqref{eq:Nu_delta_bound} we see that for $\Ra > 58.44 \dots$(actually for $\Ra > \frac{27}{4}\pi^4$)
\begin{equation}\label{eq:final_bound}
\Nu \ \leq \ \frac{5^{7/12}\cdot 3^{3/4}}{2^{14/3}} \, \Ra^{5/12} -\frac{1}{4} \ \ \lesssim \ \ 0.2295 \, \Ra^{5/12}.
\end{equation}

This $\frac{5}{12}$ exponent for the $\Nu$-$\Ra$ upper bound scaling, albeit with a prefactor $0.142$, was conjectured by Otero from a numerical study nearly a decade ago \cite{Otero2002}.  
The proof here puts that result on firm analytical ground.
The $\Nu$-$\Ra$ {\it and} the distinguished horizontal wavenumber scaling $k \sim \Ra^{1/4}$ also agree with those conjectured by Ierley, Plasting, and Kerswell following a careful combination of numerical and asymptotic analyses of the upper bound problem for infinite Prandtl number Rayleigh-B\'enard convection in {\it three} spatial dimensions with free-slip boundaries \cite{IeKePl2006}.
In fact  the analysis in this paper can be extended to that case because there is no vortex stretching at $\Pr = \infty$ so an enstrophy balance akin to \eqref{eq:energy_omega} is realized for free-slip boundaries \cite{WhDo2011b}.

While the rigorous bound $\beta \le \frac{5}{12} \approx .4167$ for the model of Rayleigh-B\'enard convection considered here is still well above that observed in most experiments and direct numerical simulations, it has significant ramifications from a theoretical point of view.
There are several theoretical predictions of $\Ra^{1/2}$ scaling of the heat transport in the ``ultimate'' regime of asymptotically high Raleigh numbers \cite{Kraichnan62,Spiegel71,GrLo2000} and the result proved here shows that those arguments cannot be correct without plainly appealing to no-slip boundary conditions or  directly relying on three dimensional dynamics (or both). 

Perhaps the simplest scaling argument---making no mention of boundaries or boundary conditions or the spatial dimension---is the hypothesis that the physical heat transport is independent of the molecular transport coefficients, i.e., the kinematic viscosity $\nu$ and the thermal diffusivity $\kappa$, in the fully developed turbulent regime \cite{Spiegel71}.
This implies $\Nu \sim \Pr^{1/2} \Ra^{1/2}$.
A more physically explicit version of the argument proceeds from the assumption that the rate-limiting process is not transferring heat across boundary layers into the bulk, but rather is the time it takes to adiabatically transport hot and cold fluid elements across the layer accelerated by the reduced gravity $\alpha  \Delta T  g$ neglecting frictional forces.
Then the vertical velocity scale of rising or falling elements is $\sqrt{g \alpha \Delta T h}$ and their heat content is ${\cal O}( \Delta T)$, so at sufficiently high density of such elements the heat flux is $\sim (g \alpha h)^{1/2} \Delta T^{3/2}$.
When normalized by the conductive heat flux $\kappa  \Delta T/h$, this again yields $\Nu \sim \Pr^{1/2} \Ra^{1/2}$.

More sophisticated arguments \cite{Kraichnan62,GrLo2000} produce the similar predictions.
It has also been proposed that the $\frac{1}{2}$ exponents will appear if the physical boundary layers are negligible (as might be hypothesized when $\Ra \rightarrow \infty$) or absent altogether.
This leads to the consideration of ``homogeneous'' Rayleigh-B\'enard convection where the Boussinesq equations with a linear background profile are posed on a fully periodic domain.
Direct numerical simulations in three dimensions and a closure theory have indicated that this scaling emerges for some aspect ratios \cite{LT2003,GOMS2010} although no upper bounds on the heat transport can possibly exist and the genuineness of statistical steady states is questionable for this formulation \cite{CDGLTT2006,GOMS2010}.

The $\Nu \lesssim \Ra^{5/12}$ bound derived here raises questions of precisely how the spatial dimension and the nature of even very thin boundary layers enter into the problem at high Rayleigh numbers.
At least in two dimensions with free-slip boundaries, no matter how high the Rayleigh number is it is apparent that boundary layers continue to play a limiting role in the turbulent heat transport.

\noindent
{\it Acknowledgements---}We thank Dr. J. Otero, Prof. J. B. Rauch, and Prof. E. A. Spiegel for helpful discussions.
This research was supported in part by NSF Award PHY-0855335.

\bibliography{references}

\end{document}